\documentclass{article}%
\usepackage{hyperref}
\usepackage{amsmath}
\usepackage{amsfonts}
\usepackage{amssymb}
\usepackage{graphicx}%
\begin{document}

\title{Finite BRST-antiBRST Transformations in Lagrangian Formalism}
\author{\textsc{Pavel Yu. Moshin${}^{a}$\thanks{moshin@rambler.ru \hspace{0.5cm}
${}^{\dagger}$reshet@ispms.tsc.ru}\ \ and Alexander A. Reshetnyak${}%
^{b,c\ddagger}$}\\\textit{${}^{a}$National Research Tomsk State University, 634050, Russia,}\\\textit{${}^{b}$Institute of Strength Physics and Materials Science, 634021,
Tomsk, Russia,}\\\textit{${}^{c}$Tomsk State Pedagogical University, 634061, Russia }}
\maketitle

\begin{abstract}
We continue the study of finite BRST-antiBRST transformations for general
gauge theories in Lagrangian formalism initiated in [arXiv:1405.0790[hep-th]],
with a doublet $\lambda_{a}$, $a=1,2$, of anticommuting Grassmann parameters,
and find an explicit Jacobian corresponding to this change of variables for
constant $\lambda_{a}$. This makes it possible to derive the Ward identities
and their consequences for the generating functional of Green's functions. We
announce the form of the Jacobian [proved to be correct in
[arXiv:1406.5086[hep-th]] for finite field-dependent BRST-antiBRST
transformations with functionally-dependent parameters, $\lambda_{a}%
=s_{a}\Lambda$, induced by a finite even-valued functional $\Lambda(\phi
,\pi,\lambda)$ and by the generators $s_{a}$ of BRST-antiBRST transformations
acting in the space of fields $\phi$, antifields $\phi_{a}^{\ast}$, $\bar
{\phi}$ and auxiliary variables $\pi^{a},\lambda$. On the basis of this
Jacobian, we solve a compensation equation for $\Lambda$, which is used to
achieve a precise change of the gauge-fixing functional for an arbitrary gauge
theory. We derive a new form of the Ward identities containing the parameters
$\lambda_{a}$ and study the problem of gauge-dependence. The general approach
is exemplified by the Freedman--Townsend model of\ a non-Abelian antisymmetric
tensor field.

\end{abstract}


\section{Introduction}

In our recent work \cite{MRnew}, we have proposed an extension of
BRST-antiBRST transformations to the case of finite (global and
field-dependent) parameters in Yang--Mills and general gauge theories within
the $\mathrm{Sp}(2)$-covariant Lagrangian quantization \cite{BLT1,BLT2}; see
also \cite{Hull}. The idea of \textquotedblleft finiteness\textquotedblright%
\ is based on transformation parameters $\lambda_{a}$ which are no longer
regarded as infinitesimal and utilizes the inclusion into the BRST-antiBRST
transformations \cite{aBRST1,aBRST2,aBRST3} of a new term, being quadratic in
$\lambda_{a}$. First of all, this makes it possible to realize the complete
BRST-antiBRST invariance of the integrand in the vacuum functional. Second,
the field-dependent parameters $\lambda_{a}=s_{a}\Lambda$, induced by a
Grassmann-even functional $\Lambda$, provide an explicit correspondence (due
to the so-called compensation equation for the Jacobian) between the partition
function of a theory in a certain gauge, determined by a gauge Boson $F_{0}$,
with the theory in a different gauge, given by another gauge Boson $F$. This
concept becomes a key instrument to determine, in a BRST-antiBRST manner, the
Gribov horizon functional \cite{Gribov} -- which is given initially in the
Landau gauge within a BRST-antiBRST extension of the Gribov--Zwanziger theory
\cite{Zwanziger} -- by utilizing any other gauge, including the $R_{\xi}%
$-gauges used to eliminate residual gauge invariance in the deep IR region.
For completeness note that concept of finite field-dependent BRST
transformations has been suggested in \cite{JM}; anti-BRST transformations and
BRST-antiBRST transformations linear in field-dependent parameters $\Theta
_{1}$, $\Theta_{2}$ have been considered in \cite{Upadhyay1} and
\cite{Upadhyay3}, respectively.

The problems listed in Discussion of \cite{MRnew} as unsolved ones include:

\begin{enumerate}
\item study of finite field-dependent BRST-antiBRST transformations for a
general gauge theory in the framework of the path integral (\ref{z(0)});

\item development of finite field-dependent BRST transformations for a general
gauge theory in the BV quantization scheme;

\item construction of finite field-dependent BRST-antiBRST transformations in
the $\mathrm{Sp}(2)$-covariant generalized Hamiltonian quantization
\cite{BLT1h, BLT2h}.
\end{enumerate}

The second problem within the BV quantization scheme \cite{BV}, based on the
principle of BRST symmetry \cite{BRST1, BRST2}, has been examined in
\cite{BLTfin}, and earlier in \cite{Reshetnyak}. The third problem has been
recently solved \cite{MRnew1} for arbitrary dynamical systems subject to
first-class constraints, together with an explicit construction of the
parameters $\lambda_{a}$ generating a change of the gauge in the path integral
for Yang--Mills theories within the class of $R_{\xi}$-like gauges in
Hamiltonian formalism. For the sake of completeness, notice that, in the case
of BRST--BFV symmetry \cite{BRST3}, a study of finite field-dependent
BRST--BFV transformations in the generalized Hamiltonian formalism \cite{BFV,
Henneaux1} has been presented in \cite{BLThf}. Therefore, it is only the first
item in the list of the above-mentioned problems that remains unsolved. In
this connection, the main purpose of the present work is to prove that the
ansatz for finite BRST-antiBRST transformations within the path integral
(\ref{z(0)}) proposed in \cite{MRnew}, using formulae (6.1)--(6.5), holds
true. We illustrate our general approach by a well-known gauge theory of
non-Yang--Mills type proposed by Freedman and Townsend \cite{FT}.

The work is organized as follows. In Section~\ref{gensetup}, we remind the
definition of a finite Lagrangian BRST-antiBRST transformation for general
gauge theories. In Section~\ref{fBRSTa}, we obtain an explicit Jacobian
corresponding to this change of variables for global finite BRST-antiBRST
transformations and prove the invariance of the integrand in the partition
function. In Section~\ref{WIGD}, we obtain the Ward identities with the help
of finite BRST-antiBRST transformations. In Section~\ref{exampleFT}, we
consider the reducible gauge theory of Freedman--Townsend (the model of
antisymmetric non-Abelian tensor field). In Discussion, we announce the
explicit Jacobian of finite field-dependent BRST-antiBRST\ transformations
with functionally-dependent parameters, formulate the corresponding
compensation equation, present its solution, which amounts to a precise change
of the gauge-fixing functional, derive the Ward identities, depending on the
parameters $\lambda_{a}$, and study the problem of gauge dependence. We use
the notation of our previous work \cite{MRnew}. In particular, derivatives
with respect to the (anti)fields are taken from the (left)right; $\delta
_{l}/\delta\phi^{A}$\ denotes the left-hand derivative with respect to
$\phi^{A}$. The raising and lowering of $\mathrm{Sp}\left(  2\right)  $
indices, $s^{a}=\varepsilon^{ab}s_{b}$, $s_{a}=\varepsilon_{ab}s^{b}$, is
carried out with the help of a constant antisymmetric tensor $\varepsilon
^{ab}$, $\varepsilon^{ac}\varepsilon_{cb}=\delta_{b}^{a}$, subject to the
normalization condition $\varepsilon^{12}=1$.

\section{Finite BRST-antiBRST Transformations}

\label{gensetup}
\renewcommand{\theequation}{\arabic{section}.\arabic{equation}}
\setcounter{equation}{0} Let $\Gamma^{p}$ be the coordinates%
\begin{equation}
\Gamma^{p}=\left(  \phi^{A},\phi_{Aa}^{\ast},\bar{\phi}_{A},\pi^{Aa}%
,\lambda^{A}\right)  \label{gencoor}%
\end{equation}
in the extended space of fields $\phi^{A}$, antifields $\phi_{Aa}^{\ast}$,
$\bar{\phi}_{A}$ and auxiliary fields $\pi^{Aa}$, $\lambda^{A}$, with the
following distribution of Grassmann parity and ghost number:%
\begin{align}
\varepsilon\left(  \phi^{A},\ \phi_{Aa}^{\ast},\ \bar{\phi}_{A},\ \pi
^{Aa},\ {\lambda}^{A}\right)   &  =\left(  \varepsilon_{A},\ \varepsilon
_{A}+1,\ \varepsilon_{A},\ \varepsilon_{A}+1,\ \varepsilon_{A}\right)
\ ,\label{Grassmann}\\
\mathrm{gh}\left(  \phi^{A},\ \phi_{Aa}^{\ast},\ \bar{\phi}_{A},\ \pi
^{Aa},\ \lambda^{A}\right)   &  =\left(  \mathrm{gh}(\phi^{A}),\ (-1)^{a}%
-\mathrm{gh}(\phi^{A}),\ -\mathrm{gh}(\phi^{A}),\ (-1)^{a+1}+\mathrm{gh}%
(\phi^{A}),\ \mathrm{gh}(\phi^{A})\right)  \ .\label{ghost}%
\end{align}
The contents of the configuration space $\phi^{A}$, containing the classical
fields $A^{i}$ and the $\mathrm{Sp}(2)$-symmetric ghost-antighost and
Nakanishi--Lautrup fields, depends on the irreducible \cite{BLT1} or reducible
\cite{BLT2} nature of a given gauge theory.

The generating functional of Green's functions $Z_{F}(J)$, depending on
external sources $J_{A}$, with $\varepsilon(J_{A})=\varepsilon_{A}$,
\textrm{gh}$(J_{A})=-\mathrm{gh}(\phi^{A})$,%
\begin{equation}
Z_{F}(J)=\int d\Gamma\;\exp\left\{  \left(  i/\hbar\right)  \left[
\mathcal{S}_{F}\left(  \Gamma\right)  +J_{A}\phi^{A}\right]  \right\}
\ ,\ \ \ \mathcal{S}_{F}=S+\phi_{Aa}^{\ast}\pi^{Aa}+\left(  \bar{\phi}%
_{A}-F_{,A}\right)  \lambda^{A}-\left(  1/2\right)  \varepsilon_{ab}\pi
^{Aa}F_{,AB}\pi^{Bb} \label{z(0)}%
\end{equation}
and the corresponding partition function $Z_{F}\equiv Z_{F}(0)$ are determined
by a Bosonic functional $S=S(\phi,\phi^{\ast},\bar{\phi})$ and by a
gauge-fixing Bosonic functional $F=F(\phi)$ with vanishing ghost numbers, the
functional $S$ being a solution of the generating equations%
\begin{equation}
\frac{1}{2}(S,S)^{a}+V^{a}S=i\hbar\Delta^{a}S\Longleftrightarrow\left(
\Delta^{a}+\frac{i}{\hbar}V^{a}\right)  \exp\left(  \frac{i}{\hbar}S\right)
=0\ , \label{3.3}%
\end{equation}
where $\hbar$ is the Planck constant, and the boundary condition for $S$ in
(\ref{3.3}) for vanishing antifields $\phi_{a}^{\ast}$, $\bar{\phi}$ is given
by the classical action $S_{0}(A)$. The extended antibracket $(F,G)^{a}$ for
arbitrary functionals $F$, $G$ and the operators $\Delta^{a}$, $V^{a}$ are
given by%
\begin{equation}
(F,G)^{a}=\frac{\delta F}{\delta\phi^{A}}\frac{\delta G}{\delta\phi_{Aa}%
^{\ast}}-\frac{\delta_{r} F}{\delta\phi_{Aa}^{\ast}}\frac{\delta_{l} G}%
{\delta\phi^{A}}\ ,\ \ \ \Delta^{a}=(-1)^{\varepsilon_{A}}\frac{\delta_{l}%
}{\delta\phi^{A}}\frac{\delta}{\delta\phi_{Aa}^{\ast}}\ ,\ \ \ V^{a}%
=\varepsilon^{ab}\phi_{Ab}^{\ast}\frac{\delta}{\delta\bar{\phi}_{A}}\ .
\label{abrack}%
\end{equation}
The integrand $\mathcal{I}_{\Gamma}^{\left(  F\right)  }=d\Gamma\exp\left[
\left(  i/\hbar\right)  \mathcal{S}_{F}\left(  \Gamma\right)  \right]  $ for
$J_{A}=0$ is invariant, $\delta\mathcal{I}_{\Gamma}^{\left(  F\right)  }=0$,
under the global infinitesimal BRST-antiBRST transformations (\ref{Gamma}),
$\delta\Gamma^{p}=\left(  s^{a}\Gamma^{p}\right)  \mu_{a}$, with the
corresponding generators $s^{a}$,%
\begin{equation}
\delta\Gamma^{p}=\left(  s^{a}\Gamma^{p}\right)  \mu_{a}=\Gamma^{p}%
\overleftarrow{s}{}^{a}\mu_{a}=\delta\left(  \phi^{A},\ \phi_{Ab}^{\ast
},\ \bar{\phi}_{A},\ \pi^{Ab},\ \lambda^{A}\right)  =\left(  \pi^{Aa}%
,\ \delta_{b}^{a}S_{,A}\left(  -1\right)  ^{\varepsilon_{A}},\ \varepsilon
^{ab}\phi_{Ab}^{\ast}\left(  -1\right)  ^{\varepsilon_{A}+1},\ \varepsilon
^{ab}\lambda^{A},\ 0\right)  \mu_{a}\ , \label{Gamma}%
\end{equation}
where the invariance at the first order in $\mu_{a}$ is established by using
the generating equations (\ref{3.3}).

The above infinitesimal invariance is sufficient to determine \emph{finite
BRST-antiBRST transformations}, $\Gamma^{p}\rightarrow\Gamma^{p}+\Delta
\Gamma^{p}$ with anticommuting parameters $\lambda_{a}$, $a=1,2$, which were
introduced in \cite{MRnew} as follows:%
\begin{equation}
\mathcal{I}_{\Gamma+\Delta\Gamma}^{\left(  {F}\right)  }=\mathcal{I}_{\Gamma
}^{\left(  {F}\right)  }\ ,\ \ \ \left.  \Delta\Gamma^{p}\frac{\overleftarrow
{\partial}}{\partial\lambda_{a}}\right\vert _{\lambda=0}=\Gamma^{p}%
\overleftarrow{s}{}^{a}\ \ \ \mathrm{and}\mathtt{\ \ \ }\Delta\Gamma^{p}%
\frac{\overleftarrow{\partial}}{\partial\lambda_{b}}\frac{\overleftarrow
{\partial}}{\partial\lambda_{a}}=\frac{1}{2}\varepsilon^{ab}\Gamma
^{p}\overleftarrow{s}{}^{2},\ \ \ \mathrm{where}\ \ \ s^{2}=s_{a}%
s^{a}\ ,\ \ \ \overleftarrow{s}{}^{2}=\overleftarrow{s}{}^{a}\overleftarrow
{s}_{a}\ . \label{Gamma_fin_def}%
\end{equation}
Thus determined finite BRST-antiBRST symmetry transformations for the
integrand $\mathcal{I}_{\Gamma}^{\left(  {F}\right)  }$ in a general gauge
theory, with the help of the notation%
\begin{equation}
{X}^{pa}\equiv\Gamma^{p}\overleftarrow{s}{}^{a}\ \ \ \mathrm{and}\ \ \ {Y}%
^{p}\equiv\left(  1/2\right)  {X}_{,q}^{pa}{X}^{qb}\varepsilon_{ba}=-\left(
1/2\right)  \Gamma^{p}\overleftarrow{s}{}^{2}\ ,\ \ \ \mathrm{with}%
\ \ \ G_{,p}\equiv\frac{\delta G}{\delta\Gamma^{p}}\ , \label{auxcomp}%
\end{equation}
can be represented in the form%
\begin{equation}
\Delta\Gamma^{p}={X}^{pa}\lambda_{a}-\frac{1}{2}{Y}^{p}\lambda^{2}=\Gamma
^{p}\left(  \overleftarrow{s}{}^{a}\lambda_{a}+\frac{1}{4}\overleftarrow{s}%
{}^{2}\lambda^{2}\right)  \Longrightarrow\mathcal{I}_{\Gamma+\Delta\Gamma
}^{\left(  _{F}\right)  }=\mathcal{I}_{\Gamma}^{\left(  _{F}\right)  }\ .
\label{Gamma_fin}%
\end{equation}
Equivalently, in terms of the components, (\ref{Gamma_fin}) is given by%
\begin{align}
\Delta\phi^{A}  &  =\pi^{Aa}\lambda_{a}+\frac{1}{2}\lambda^{A}\lambda
^{2}%
,\phantom{= \pi^{Aa}\lambda_{a}+ \frac{1}{2}\lambda^{A} \lambda^{2}\quad\quad\ }\Delta
\bar{\phi}_{A}\ =\ \varepsilon^{ab}\lambda_{a}\phi_{Ab}^{\ast}+\frac{1}%
{2}S_{,A}\lambda^{2}\ ,\nonumber\\
\Delta\pi^{Aa}  &  =-\varepsilon^{ab}\lambda^{A}\lambda_{b}%
\ ,\phantom{= \pi^{Aa}\lambda_{a}+ \frac{1}{2}\lambda^{A} \lambda^{2}=\varepsilon^{ab}\lambda^{A}\lambda_{a}}\Delta
\lambda^{A}\ =\ 0\ ,\label{exply}\\
\Delta\phi_{Aa}^{\ast}  &  =\lambda_{a}S_{,A}+\frac{1}{4}\left(  -1\right)
^{\varepsilon_{A}}\left(  \varepsilon_{ab}\frac{\delta^{2}S}{\delta\phi
^{A}\delta\phi^{B}}\pi^{Bb}+\varepsilon_{ab}\frac{\delta S}{\delta\phi^{B}%
}\frac{\delta^{2}S}{\delta\phi^{A}\delta\phi_{Bb}^{\ast}}\left(  -1\right)
^{\varepsilon_{B}}-\phi_{Ba}^{\ast}\frac{\delta^{2}S}{\delta\phi^{A}\delta
\bar{\phi}_{B}}\left(  -1\right)  ^{\varepsilon_{B}}\right)  \lambda
^{2}\ .\nonumber
\end{align}
In order to make sure that $\mathcal{I}_{\Gamma}^{\left(  {F}\right)  }$ is
invariant under the finite BRST-antiBRST transformations (\ref{Gamma_fin})
with constant $\lambda_{a}$, one has to find the Jacobian corresponding to
this change of variables.

\section{Jacobian of Finite Global BRST-antiBRST Transformations}

\label{fBRSTa} \renewcommand{\theequation}{\arabic{section}.\arabic{equation}} \setcounter{equation}{0}

Let us examine the change of the integration measure $d\Gamma\rightarrow
d\check{\Gamma}$ in (\ref{z(0)}) under the finite transformations $\Gamma
^{p}\rightarrow\check{\Gamma}^{p}=\Gamma^{p}+\Delta\Gamma^{p}$ given by
(\ref{Gamma_fin}). To this end, taking account of (\ref{3.3}), we present the
invariance of the integrand $\mathcal{I}_{\Gamma}^{\left(  _{F}\right)  }$
under the infinitesimal transformations $\delta\Gamma^{p}=\Gamma
^{p}\overleftarrow{s}{}^{a}\mu_{a}=X_{a}^{pa}\mu_{a}$ given by (\ref{Gamma})
in the form%
\begin{equation}
\mathcal{S}_{F,p}X^{pa}=i\hbar X_{,p}^{pa}\ ,\ \ \mathrm{where}\ \ \ X_{_{,}%
p}^{pa}=-\Delta^{a}S\ . \label{eqinv}%
\end{equation}
Considering (\ref{Gamma_fin}) implies that we are interested in%
\begin{equation}
\mathrm{Str}\left(  M-\frac{1}{2}M^{2}\right)  \ ,\ \ \mathrm{for}%
\ \ M_{q}^{p}\equiv\frac{\delta\left(  \Delta\Gamma^{p}\right)  }{\delta
\Gamma^{q}}\ \ \ \mathrm{ with }\ \ \frac{\delta}{\delta\Gamma^{q}}
\equiv\frac{\delta_{r} }{\delta\Gamma^{q}}, \label{Str}%
\end{equation}
since, in view of the nilpotency $\lambda_{a}\lambda_{b}\lambda_{c}\equiv0$,
we have%
\begin{align*}
&  d\check{\Gamma}=d\Gamma\ \mathrm{Sdet}\left(  \frac{\delta\check{\Gamma}%
}{\delta\Gamma}\right)  =d\Gamma\ \exp\left[  \mathrm{Str\ln}\left(
\mathbb{I+}M\right)  \right]  \equiv d\Gamma\ \exp\left(  \Im\right)  \ ,\\
&  \ \Im= \mathrm{Str\ln}\left(  \mathbb{I+}M\right)  =-\mathrm{Str}\left(
\sum_{n=1}^{\infty}\frac{\left(  -1\right)  ^{n}}{n}M^{n}\right)  =
\mathrm{Str}\left(  M-\frac{1}{2}M^{2}\right) \ .
\end{align*}
Explicitly,%
\begin{align}
&  M_{q}^{p}=\frac{\delta\left(  \Delta\Gamma^{p}\right)  }{\delta\Gamma^{q}%
}=\frac{\delta}{\delta\Gamma^{q}}\left(  X^{pa}\lambda_{a}-\frac{1}{2}%
Y^{p}\lambda^{2}\right)  =\left(  -1\right)  ^{\varepsilon_{q}}X_{,q}%
^{pa}\lambda_{a}-\frac{1}{2}Y_{,q}^{p}\lambda^{2}\ ,\nonumber\\
&  \ \mathrm{with}\ \ \mathrm{Str}\left(  M\right)  =X_{,p}^{pa}\lambda
_{a}-\frac{1}{2}\left(  -1\right)  ^{\varepsilon_{p}}Y_{,p}^{p}\lambda^{2}
\label{strmpq}%
\end{align}
and%
\begin{align}
&  M_{r}^{p}M_{q}^{r}=\left(  -1\right)  ^{\varepsilon_{r}}X_{,r}^{pa}%
\lambda_{a}\left(  -1\right)  ^{\varepsilon_{q}}X_{,q}^{rb}\lambda_{b}%
=X_{,r}^{pa}X_{,q}^{rb}\lambda_{b}\lambda_{a}=-\frac{1}{2}\varepsilon
_{ba}X_{,r}^{pa}X_{,q}^{rb}\lambda^{2}\ ,\nonumber\\
&  \ \mathrm{with}\ \ \mathrm{Str}\left(  M^{2}\right)  =-\frac{1}{2}\left(
-1\right)  ^{\varepsilon_{p}}X_{,q}^{pa}X_{,p}^{qb}\varepsilon_{ba}\lambda
^{2}\ . \label{strmpq2}%
\end{align}
Therefore,%
\begin{align}
\mathrm{Str}\left(  M-\frac{1}{2}M^{2}\right)   &  =X_{,p}^{pa}\lambda
_{a}-\frac{1}{2}\left(  -1\right)  ^{\varepsilon_{p}}Y_{,p}^{p}\lambda
^{2}-\frac{1}{2}\left(  -\frac{1}{2}\left(  -1\right)  ^{\varepsilon_{p}%
}X_{,q}^{pa}X_{,p}^{qb}\varepsilon_{ba}\lambda^{2}\right) \nonumber\\
&  =X_{,p}^{pa}\lambda_{a}-\frac{1}{2}\left(  -1\right)  ^{\varepsilon_{p}%
}Y_{,p}^{p}\lambda^{2}+\frac{1}{4}\left(  -1\right)  ^{\varepsilon_{p}}%
X_{,q}^{pa}X_{,p}^{qb}\varepsilon_{ba}\lambda^{2}\nonumber\\
&  =X_{,p}^{pa}\lambda_{a}-\frac{1}{2}\left(  -1\right)  ^{\varepsilon_{p}%
}\left(  Y_{,p}^{p}-\frac{1}{2}X_{,q}^{pa}X_{,p}^{qb}\varepsilon_{ba}\right)
\lambda^{2}\ . \label{strm-m2}%
\end{align}
Considering%
\begin{align}
Y_{,p}^{p}-\frac{1}{2}X_{,q}^{pa}X_{,p}^{qb}\varepsilon_{ba}  &  =\frac{1}%
{2}\varepsilon_{ba}\left(  X_{,qp}^{pa}X^{qb}\left(  -1\right)  ^{\varepsilon
_{p}\left(  \varepsilon_{q}+1\right)  }+X_{,q}^{pa}X_{,p}^{qb}\right)
-\frac{1}{2}\varepsilon_{ba}X_{,q}^{pa}X_{,p}^{qb}\nonumber\\
&  =\frac{1}{2}\varepsilon_{ba}\left(  X_{,qp}^{pa}X^{qb}\left(  -1\right)
^{\varepsilon_{p}\left(  \varepsilon_{q}+1\right)  }+X_{,q}^{pa}X_{,p}%
^{qb}-X_{,q}^{pa}X_{,p}^{qb}\right)  =\frac{1}{2}\varepsilon_{ba}X_{,pq}%
^{pa}X^{qb}\left(  -1\right)  ^{\varepsilon_{p}}\ , \label{consstrmpq}%
\end{align}
we arrive at%
\begin{equation}
\mathrm{Str}\left(  M-\frac{1}{2}M^{2}\right)  =X_{,p}^{pa}\lambda_{a}%
+\frac{1}{4}\varepsilon_{ab}X_{,pq}^{pa}X^{qb}\lambda^{2}\ , \label{lnjacob}%
\end{equation}
where (\ref{eqinv}) implies%
\begin{equation}
X_{,p}^{pa}=-\Delta^{a}S\ ,\ \ \ X_{,pq}^{pa}X^{qb}=-\left(  \Delta
^{a}S\right)  _{,p}X^{pb}=-s^{b}\left(  \Delta^{a}S\right)
\ ,\ \ \mathrm{with}\ \ G_{,p}X^{pa}=G_{,p}\left(  s^{a}\Gamma^{p}\right)
=s^{a}G\ . \label{diffcons}%
\end{equation}
Hence, (\ref{lnjacob}) takes the form%
\begin{equation}
\mathrm{Str}\left(  M-\frac{1}{2}M^{2}\right)  =-\left(  \Delta^{a}S\right)
\lambda_{a}-\frac{1}{4}\varepsilon_{ab}\left(  \Delta^{a}S\right)  _{,p}%
X^{pb}\lambda^{2}=-\left(  \Delta^{a}S\right)  \lambda_{a}-\frac{1}{4}\left(
s_{a}\Delta^{a}S\right)  \lambda^{2}\ . \label{lnjacob2}%
\end{equation}
Consider now the change of the integrand%
\begin{equation}
\mathcal{I}_{\Gamma}\equiv\mathcal{I}_{\Gamma}^{\left(  F\right)  }%
=d\Gamma\exp\left[  \left(  i/\hbar\right)  \mathcal{S}_{F}\left(
\Gamma\right)  \right]  \label{integrand}%
\end{equation}
under the transformations (\ref{Gamma_fin}),%
\begin{align}
&  \mathcal{I}_{\Gamma+\Delta\Gamma}=d\Gamma\ \mathrm{Sdet}\left(
\frac{\delta\check{\Gamma}}{\delta\Gamma}\right)  \exp\left[  \frac{i}{\hbar
}\mathcal{S}_{F}\left(  \Gamma+\Delta\Gamma\right)  \right]  \ ,\nonumber\\
&  \mathrm{Sdet}\left(  \frac{\delta\check{\Gamma}}{\delta\Gamma}\right)
=\exp\left\{  \frac{i}{\hbar}\left[  -i\hbar\ \mathrm{Str}\left(  M-\frac
{1}{2}M^{2}\right)  \right]  \right\} \nonumber\\
&  \ =\exp\left\{  \frac{i}{\hbar}\left[  i\hbar\Delta^{a}S\lambda_{a}%
+\frac{i\hbar}{4}\left(  s_{a}\Delta^{a}S\right)  \lambda^{2}\right]
\right\}  \ ,\label{sdetgl}\\
&  \ \mathcal{S}_{F}\left(  \Gamma+\Delta\Gamma\right)  \ =\ \mathcal{S}%
_{F}\left(  \Gamma\right)  +s^{a}\mathcal{S}_{F}\left(  \Gamma\right)
\lambda_{a}+\frac{1}{4}s^{2}\mathcal{S}_{F}\left(  \Gamma\right)  \lambda
^{2}\ , \label{actiongl}%
\end{align}
where any functional $G\left(  \Gamma\right)  $ expandable as a power series
in $\Gamma^{p}$,%
\[
G\left(  \Gamma+\Delta\Gamma\right)  =G\left(  \Gamma\right)  +G_{,p}\left(
\Gamma\right)  \Delta\Gamma^{p}+\left(  1/2\right)  G_{,pq}\left(
\Gamma\right)  \Delta\Gamma^{q}\Delta\Gamma^{p}\equiv G\left(  \Gamma\right)
+\Delta G\left(  \Gamma\right)  \ ,
\]
transforms under (\ref{Gamma_fin}) as%
\begin{align}
\Delta G  &  =G_{,p}X^{pa}\lambda_{a}-\frac{1}{2}G_{,p}Y^{p}\lambda^{2}%
+\frac{1}{2}G_{,pq}X^{qb}\lambda_{b}X^{pa}\lambda_{a}\nonumber\\
&  =\left(  G_{,p}X^{pa}\right)  \lambda_{a}+\frac{1}{2}\left(  \frac{1}%
{2}\varepsilon_{ab}G_{,qp}X^{pa}X^{qb}\left(  -1\right)  ^{\varepsilon_{q}%
}-G_{,p}Y^{p}\right)  \lambda^{2}=\left(  s^{a}G\right)  \lambda_{a}+\frac
{1}{4}\left(  s^{2}G\right)  \lambda^{2}\ . \label{DeltaF}%
\end{align}
From (\ref{sdetgl}), (\ref{actiongl}), it follows that%
\begin{align}
\mathcal{I}_{\Gamma+\Delta\Gamma}  &  =d\Gamma\ \exp\left\{  \frac{i}{\hbar
}\left[  i\hbar\left(  \Delta^{a}S\right)  \lambda_{a}+\frac{i\hbar}{4}\left(
s_{a}\Delta^{a}S\right)  \lambda^{2}\right]  \right\}  \exp\left\{  \frac
{i}{\hbar}\left[  \mathcal{S}_{F}+\left(  s^{a}\mathcal{S}_{F}\right)
\lambda_{a}+\frac{1}{4}\left(  s^{2}\mathcal{S}_{F}\right)  \lambda
^{2}\right]  \right\} \nonumber\\
&  =d\Gamma\ \exp\left(  \frac{i}{\hbar}\mathcal{S}_{F}\right)  \exp\left[
\frac{i}{\hbar}\left(  s^{a}\mathcal{S}_{F}+i\hbar\Delta^{a}S\right)
\lambda_{a}+\frac{i}{4\hbar}s_{a}\left(  s^{a}\mathcal{S}_{F}+i\hbar\Delta
^{a}S\right)  \lambda^{2}\right] \nonumber\\
&  =d\Gamma\exp\left(  \frac{i}{\hbar}\mathcal{S}_{F}\right)  =\mathcal{I}%
_{\Gamma}\ , \label{finident}%
\end{align}
since\textrm{ }$s^{a}\mathcal{S}_{F}+i\hbar\Delta^{a}S=0$, due to
(\ref{eqinv}),$\ $which proves that the change of variables $\Gamma
^{p}\rightarrow\Gamma^{p}+\Delta\Gamma^{p}$ in (\ref{Gamma_fin}) realizes
finite BRST-antiBRST transformations. By virtue of (\ref{lnjacob2}), the
Jacobian of finite BRST-antiBRST transformations (\ref{Gamma_fin}) with
constants parameters $\lambda_{a}$ equals to%
\begin{equation}
\exp\left(  \Im\right)  =\exp\left[  -\left(  \Delta^{a}S\right)  \lambda
_{a}-\frac{1}{4}\left(  \Delta^{a}S\right)  \overleftarrow{s}_{a}\lambda
^{2}\right]  \ . \label{ansjacob}%
\end{equation}

\section{Ward Identities}

\label{WIGD}\renewcommand{\theequation}{\arabic{section}.\arabic{equation}}
\setcounter{equation}{0} We can now apply the finite global BRST-antiBRST
transformations to obtain the Ward (Slavnov--Taylor) identities for the
generating functional of Green's functions (\ref{z(0)}). Namely, using the
Jacobian (\ref{ansjacob}) of finite BRST-antiBRST transformations with
constants parameters $\lambda_{a}$, we make a change of variables
(\ref{Gamma_fin}) in the integrand (\ref{z(0)}) for $Z_{F}(J)$ and arrive at%
\begin{equation}
\left\langle \left[  1+\frac{i}{\hbar}J_{A}\phi^{A}\left(  \overleftarrow{s}%
{}^{a}\lambda_{a}+\frac{1}{4}\overleftarrow{s}{}^{2}\lambda^{2}\right)
-\frac{1}{4}\left(  \frac{i}{\hbar}\right)  {}^{2}J_{A}\phi^{A}\overleftarrow
{s}{}^{a}J_{B}(\phi^{B})\overleftarrow{s}_{a}\lambda^{2}\right]  \right\rangle
_{F,J}=1\ . \label{WIsp2}%
\end{equation}
Here, the symbol \textquotedblleft$\langle\mathcal{O}\rangle_{F,J}%
$\textquotedblright\ for a quantity $\mathcal{O}=\mathcal{O}(\Gamma)$ stands
for the source-dependent average expectation value corresponding to a
gauge-fixing $F(\phi)$, namely,%
\begin{equation}
\left\langle \mathcal{O}\right\rangle _{F,J}=Z_{F}^{-1}(J)\int d\Gamma
\ \mathcal{O}\left(  \Gamma\right)  \exp\left\{  \frac{i}{\hbar}\left[
\mathcal{S}_{F}\left(  \Gamma\right)  +J_{A}\phi^{A}\right]  \right\}
\ ,\ \ \mathrm{with\ \ }\left\langle 1\right\rangle _{F,J}=1\ . \label{aexv}%
\end{equation}
The relation (\ref{WIsp2}) is the Ward identity, depending on a doublet of
arbitrary constants $\lambda_{a}$ and on sources $J_{A}$. Using an expansion
in powers of $\lambda_{a}$, we obtain, at the first order, the usual Ward
identities%
\begin{equation}
J_{A}\left\langle \phi^{A}\overleftarrow{s}{}^{a}\right\rangle _{F,J}=0
\label{WIlag1}%
\end{equation}
and a new Ward identity, at the second order:%
\begin{equation}
\left\langle J_{A}\phi^{A}\left[  \overleftarrow{s}{}^{2}-\overleftarrow
{s}^{a}\left(  i/\hbar\right)  J_{B}\left(  \phi^{B}\overleftarrow{s}%
_{a}\right)  \right]  \right\rangle _{F,J}=0\ . \label{WIlag2}%
\end{equation}

\section{Freedman--Townsend Model}

\label{exampleFT}%
\renewcommand{\theequation}{\arabic{section}.\arabic{equation}} \setcounter{equation}{0}

In this section, we illustrate the above construction of finite BRST-antiBRST
transformations in general gauge theories by using the example of a well-known
theory of non-Yang-Mills type, being the reducible gauge model \cite{FT}
suggested by Freedman and Townsend, whose Lagrangian quantization and
investigation of the unitarity problem have been considered in the BRST
\cite{SFr, LTy} and BRST-antiBRST \cite{lm-ft, lm-phun} symmetries. To this
end, let us consider the theory of a non-Abelian antisymmetric tensor field
$\mathfrak{B}_{\mu\nu}^{m}$ given in Minkowski space $\mathbb{R}^{1,3}$ by the
action \cite{FT}%
\begin{equation}
S_{0}(A,\mathfrak{B})=\int{d^{4}}x\left(  -\frac{1}{4}{\varepsilon}^{\mu
\nu\rho\sigma}F_{\mu\nu}^{m}\mathfrak{B}_{\rho\sigma}^{m}+\frac{1}{2}A_{\mu
}^{m}A^{m\mu}\right)  \ ,\label{5.1}%
\end{equation}
with the Lorentz indices $\mu,\nu\,\rho,\sigma=0,1,2,3$, the metric tensor
$\eta_{\mu\nu}=\mathrm{diag}(-,+,+,+)$, the completely antisymmetric structure
constants $f^{lmn}$ of the Lie algebra $su(N)$ for $l,m,n=1,\ldots,N^{2}-1$;
$A_{\mu}^{m}$ is a vector gauge field with the strength $F_{\mu\nu}^{m}%
\equiv{\partial}_{\mu}A_{\nu}^{m}-{\partial}_{\nu}A_{\mu}^{m}+f^{mnl}A_{\mu
}^{n}A_{\nu}^{l}$ (the coupling constant is absorbed into the structure
coefficients $f^{mnl}$), and ${\varepsilon}^{\mu\nu\rho\sigma}$ is a constant
completely antisymmetric four-rank tensor, ${\varepsilon}^{0123}=1$. The
action (\ref{5.1}) is invariant under the gauge transformations%
\begin{equation}
{\delta}\mathfrak{B}_{\mu\nu}^{m}=D_{\mu}^{mn}{\zeta}_{\nu}^{n}-D_{\nu\mu
}^{mn}{\zeta}^{n}\equiv R_{\mu\nu\rho}^{mn}{\zeta}^{n\rho}%
\ ,\,\,\,\,\,\,\,\,{\delta}A_{\mu}^{m}=0\ ,\ \ \ \mathrm{for}\ \ \ D_{\mu
}^{mn}={\delta}^{mn}{\partial}_{\mu}+f^{mln}A_{\mu}^{l}\ ,\label{5.2}%
\end{equation}
where $\zeta_{\mu}^{m}$ are arbitrary Bosonic functions, and $D_{\mu}^{mn}$ is
the covariant derivative with potential $A_{\mu}^{m}$. The algebra of the
gauge transformations (\ref{5.2}) is Abelian, and the generators $R_{\mu
\nu\rho}^{mn}$ have at the extremals of the action (\ref{5.1}) the Bosonic
zero-eigenvectors $Z_{\mu}^{mn}\equiv D_{\mu}^{mn}$,%
\begin{equation}
R_{\mu\nu\rho}^{ml}Z^{ln\rho}={\varepsilon}_{\mu\nu\rho\sigma}f^{mln}%
\frac{\delta S_{0}}{\delta\mathfrak{B}_{\rho\sigma}^{l}}\,,\label{5.3}%
\end{equation}
which are linearly independent. By the generally accepted terminology
\cite{BV}, the model (\ref{5.1})--(\ref{5.3}) is an Abelian gauge theory of
first-stage reducibility. In accordance with the Lagrangian $\mathrm{Sp}%
(2)$-symmetric quantization \cite{BLT2} for reducible gauge theories, the
fields $\phi^{A}$ and the corresponding antifields ${\phi}_{Aa}^{\ast}$,
$\bar{\phi}_{A}$ for the model (\ref{5.1})--(\ref{5.3}) are given by%
\begin{align}
&  {\phi}^{A}=(A^{m\mu};\mathfrak{B}^{m\mu\nu},B^{m\mu},B^{ma},C^{m\mu
a},C^{mab})\ ,\nonumber\\
&  \phi_{Aa}^{\ast}=(A_{\mu a}^{m\ast};\mathfrak{B}_{\mu\nu a}^{m\ast},B_{\mu
a}^{m\ast},B_{a|b}^{m\ast},C_{\mu a|b}^{m\ast},C_{a|bc}^{m\ast})\ ,\ \ \ \bar
{\phi}_{A}=(\bar{A}_{\mu}^{m};\bar{\mathfrak{B}}_{\mu\nu}^{m},\bar{B}_{\mu
}^{m},\bar{B}_{a}^{m},\bar{C}_{\mu a}^{m},\bar{C}_{ab}^{m})\ ,\label{totspace}%
\end{align}
where $B^{ma}$ and $C^{mab}$ are the respective $\mathrm{Sp}(2)$-doublets of
fields introducing the gauge and the ghost fields (symmetric second rank
$\mathrm{Sp}(2)$-tensors) of the first stage, in accordance with the number of
gauge parameters $\zeta^{m}$ for the generators $R_{1\mu\nu}^{mn}\equiv
R_{\mu\nu\rho}^{ml}Z^{ln\rho}$. Taking account of (\ref{Grassmann}),
(\ref{ghost}), the Grassmann parity and ghost number of the variables
($\phi^{A}$, ${\phi}_{Aa}^{\ast}$, $\bar{\phi}_{A}$) are given by%
\begin{align}
&  \varepsilon\left(  A^{m\mu};\mathfrak{B}^{m\mu\nu},B^{m\mu},B^{ma},C^{m\mu
a},C^{mab}\right)  =\left(  0;0,0,1,1,0\right)  \ ,\label{Grassmann1}\\
&  \mathrm{gh}\left(  A^{m\mu};\mathfrak{B}^{m\mu\nu},B^{m\mu},B^{ma},C^{m\mu
a},C^{mab}\right)  =\left(  0;0,0,3-2a,3-2a,6-2(a+b)\right)  \ .\label{ghost1}%
\end{align}
A solution $S=S(\phi,\phi^{\ast},\bar{\phi})$ of the generating equations
(\ref{3.3}) with the boundary condition $\left.  S\right\vert _{\phi^{\ast
}=\bar{\phi}=0}=S_{0}$ for the model (\ref{5.1})--(\ref{5.3}) can be
represented in the form being quadratic in powers of the antifields,%
\begin{align}
S &  =S_{0}+\int d^{4}x\,\left[  \mathfrak{B}_{\mu\nu a}^{\ast}\left(  D^{\mu
}C^{\nu a}-D^{\nu}C^{\mu a}-\varepsilon^{\mu\nu\rho\sigma}\bar{\mathfrak{B}%
}_{\rho\sigma}\wedge B^{a}\right)  -\varepsilon^{ab}C_{\mu a|b}^{\ast}B^{\mu
}+\bar{\mathfrak{B}}_{\mu\nu}(D^{\mu}B^{\nu}-D^{\nu}B^{\mu})\right.
\nonumber\\
&  \left.  +C_{\mu a|b}^{\ast}D^{\mu}C^{ab}-2\varepsilon^{ab}C_{a|bc}^{\ast
}B^{c}-B_{\mu a}^{\ast}D^{\mu}B^{a}+2\bar{C}_{\mu a}D^{\mu}B^{a}+\frac{1}%
{2}\varepsilon^{\mu\nu\rho\sigma}(\mathfrak{B}_{\mu\nu a}^{\ast}%
\wedge\mathfrak{B}_{\rho\sigma b}^{\ast})C^{ab}\right]  \ ,\label{S(FT)}%
\end{align}
with the following notation for the fields $A^{m}\equiv A$, $B^{m}\equiv B$:%
\begin{equation}
A^{m}B^{m}\equiv AB\ ,\ \ \ D_{\mu}B\equiv\partial_{\mu}B+A_{\mu}\wedge
B\ ,\;\;(A\wedge B)^{m}=f^{mnl}A^{n}B^{l}\ .\label{auxFT}%
\end{equation}
Choosing the gauge Boson $F=F(\phi)$ in the form of a $3$-parametric quadratic
functional,%
\begin{equation}
F(\alpha,\beta,\gamma)=\int d^{4}x\,\left(  -\frac{\alpha}{4}\mathfrak{B}%
_{\mu\nu}\mathfrak{B}^{\mu\nu}-\frac{\beta}{2}\varepsilon_{ab}C_{\mu}%
^{a}C^{\mu b}-\frac{\gamma}{12}\varepsilon_{ab}\varepsilon_{cd}C^{ac}%
C^{bd}\right)  ,\ \ \mathrm{for}\ \ \alpha,\beta,\gamma\in\mathbb{R\ }%
,\label{F(FT)}%
\end{equation}
and integrating in (\ref{z(0)}) over the variables $\lambda$, $\pi^{a}$,
$\bar{\phi}$, $\phi_{a}^{\ast}$, we obtain the generating functional of
Green's functions%
\begin{equation}
Z_{F}(J)=\int d\phi\;\Delta_{\alpha}\left(  \phi\right)  \;\exp\left\{
\left(  i/\hbar\right)  \left[  S_{0}\left(  A\right)  +S_{\mathrm{gf}}\left(
\phi\right)  +S_{\mathrm{fp}}\left(  \phi\right)  +J_{A}\phi^{A}\right]
\right\}  \ ,\label{z(FT)}%
\end{equation}
indentical with that of \cite{lm-phun} in the case $(\alpha,\beta
,\gamma)=(\alpha_{0},\beta_{0},\gamma_{0})\equiv(1,2,1)$, corresponding to
$F_{0}\equiv F(1,2,1)$, where%
\begin{align}
S_{\mathrm{gf}} &  =\int d^{4}x\left(  \alpha B_{\mu}D_{\nu}\mathfrak{B}%
^{\nu\mu}+\beta\varepsilon_{ab}B^{a}D_{\mu}C^{\mu b}-\beta B_{\mu}B^{\mu
}-\frac{\gamma}{2}\varepsilon_{ab}B^{a}B^{b}\right)  \ ,\label{gh}\\
S_{\mathrm{fp}} &  =\int d^{4}x\left(  \frac{\alpha}{4}G_{\mu\nu}^{a}%
M_{ab}K_{c}^{b[\mu\nu][\rho\sigma]}G_{\rho\sigma}^{c}-\frac{\beta}%
{2}\varepsilon_{ab}\varepsilon_{cd}D_{\mu}C^{ac}D^{\mu}C^{bd}\right)
\ ,\label{fp}\\
\Delta_{\alpha} &  =\int d\mathfrak{B}^{\ast}\ \exp\left(  \frac{2i}%
{\alpha\hbar}\int d^{4}x\,\mathfrak{B}_{0ib}^{\ast}M^{bc}\mathfrak{B}%
_{0jc}^{\ast}\eta^{ij}\right)  \ .\label{delta}%
\end{align}
In (\ref{fp}), (\ref{delta}) we have used the notation%
\begin{equation}
K_{b}^{a[\mu\nu][\rho\sigma]}\equiv\frac{1}{2}\left[  \delta_{b}^{a}(\eta
^{\mu\rho}\eta^{\nu\sigma}-\eta^{\mu\sigma}\eta^{\nu\rho})+\alpha X_{b}%
^{a}\varepsilon^{\mu\nu\rho\sigma}\right]  \ ,\ \ \ G_{\mu\nu}^{a}\equiv
D_{\mu}C_{\nu}^{a}-D_{\nu}C_{\mu}^{a}-\frac{\alpha}{4}\varepsilon_{\mu\nu
\rho\sigma}Y^{a}\mathfrak{B}^{\rho\sigma}\ ,\label{notK}%
\end{equation}
and the matrix $M_{ab}$ is the inverse of $M^{ab}$,%
\begin{equation}
M^{ab}\equiv\varepsilon^{ab}-\alpha^{2}X_{c}^{a}X_{d}^{b}\varepsilon
^{cd}\ ,\ \ \ M^{ac}M_{cb}=\delta_{b}^{a}\ ,\label{Mab}%
\end{equation}
while the action of the matricrs $X_{b}^{a}$ and $Y^{a}$ on the objects
$E\equiv E^{m}$ carrying the indices $m$ is given by the rule%
\begin{equation}
X_{b}^{a}E\equiv\varepsilon_{bc}(C^{ac}\wedge E)\ ,\ \ \ Y^{a}E\equiv
(B^{a}\wedge E)=-(-1)^{\varepsilon(E)}EY^{a}\ .\label{Xba}%
\end{equation}
For the vanishing sources, $J=0$, the integrand in (\ref{z(FT)}) is invariant
under the BRST-antiBRST transformations \cite{lm-ft} in the space of fields
$\phi^{A}$%
\begin{equation}%
\begin{array}
[c]{lll}%
\delta\mathfrak{B}^{\mu\nu}=-\varepsilon^{ab}M_{bc}K_{d}^{c[\mu\nu][\rho
\sigma]}G_{\rho\sigma}^{d}\mu_{a}\ , & \delta A^{\mu}=0\ , & \delta C^{\mu
a}=(D^{\mu}C^{ab}-\varepsilon^{ab}B^{\mu})\mu_{b}\ ,\\
\delta B^{\mu}=D^{\mu}B^{a}\mu_{a}\ , & \delta C^{ab}=B^{\{a}\varepsilon
^{b\}c}\mu_{c}\ , & \delta B^{a}=0\ .
\end{array}
\label{brst-anti-brst-FT}%
\end{equation}
Indeed, the quantum action and the integration measure under the change of
variables $\phi^{A}\rightarrow\check{\phi}{}^{A}=\phi^{A}+\delta\phi^{A}$ are
transformed as%
\begin{align}
& \delta\left(  S_{0}+S_{\mathrm{gf}}+S_{\mathrm{fp}}\right)  =0\ ,\ \ d\check
{\phi}\Delta_{\alpha}(\check{\phi})=d{\phi}\Delta_{\alpha}(\check{\phi
})\mathrm{Sdet}\left(  \delta\check{\phi}/{\delta\phi}\right)  =d{\phi}%
\Delta_{\alpha}+\delta(d\phi)\Delta_{\alpha}+d\phi\delta(\Delta_{\alpha
})=d\phi\Delta_{\alpha}(\phi),\nonumber\\
& \ \mathrm{where}\ \ \,\delta(d\phi)=\delta^{4}(0)\int d^{4}x\,\mathrm{Tr\ }%
W\ \ \ \mathrm{and}\ \ \ \delta\Delta_{\alpha}=-\Delta_{\alpha}\delta
^{4}(0)\int d^{4}x\,\mathrm{Tr\ }W\ ,\label{babft}%
\end{align}
where $\delta^{4}(0)\equiv\left.  \delta(x-y)\right\vert _{x=y}$ and we use
the notation%
\begin{equation}
W\equiv W^{mn}=-3\alpha^{2}\varepsilon^{ab}M_{bc}X_{d}^{c}Y^{d}\mu
_{a}\ ,\ \ \ \mathrm{for}\ \ \ \mathrm{Tr\ }W\equiv\sum_{m=1}^{N^{2}-1}%
W^{mm}\ .\label{not2}%
\end{equation}
The functional $\Delta_{\alpha}$ in (\ref{delta}) is a contribution to the
integration measure $d\phi\Delta_{\alpha}$, being invariant, $\delta
(d\phi\Delta_{\alpha})=0$, under the BRST-antiBRST transformations
(\ref{brst-anti-brst-FT}). At the same time, we notice that these
transformations depend explicitly on the parameter $\alpha$ of the gauge Boson
$F$ in (\ref{F(FT)}). Due to a non-trivial integration measure and
BRST-antiBRST transformations depending on a choice of the\ gauge Boson, the
task of connecting (by finite BRST-antiBRST transformations) the generating
functionals $Z_{F}(J)$ and $Z_{F+\Delta F}(J)$ given by different gauges $F$
and $F+\Delta F$ in the representation (\ref{z(FT)}) cannot be solved
literally on the basis of our approach \cite{MRnew}, developed on the basis of
a compensation equation for Yang--Mills type theories, and deserves a special
analysis \cite{newstudy}. In this connection, we restrict the conisderation to
the quantum theory (\ref{S(FT)}), (\ref{F(FT)}), with the generating
functional $Z_{F}(J)$ given by the functional integral (\ref{z(0)}) in the
extended space $\phi$, $\phi_{a}^{\ast}$, $\bar{\phi}$, $\pi^{a}$, $\lambda$,
where (omitting the $su(N)$ indices $m$)%
\begin{equation}
{\pi}^{Aa}=(\pi_{\left(  A\right)  }^{\mu a};\pi_{\left(  \mathfrak{B}\right)
}^{\mu\nu a},\pi_{\left(  B\right)  }^{\mu a},\pi_{\left(  B\right)  }%
^{a|b},\pi_{\left(  C\right)  }^{\mu a|b},\pi_{\left(  C\right)  }%
^{a|bc})\ ,\ \ \ \lambda^{A}=(\lambda_{\left(  A\right)  }^{\mu}%
;\lambda_{\left(  \mathfrak{B}\right)  }^{\mu\nu},\lambda_{\left(  B\right)
}^{\mu},\lambda_{\left(  B\right)  }^{a},\lambda_{\left(  C\right)  }^{\mu
a},\lambda_{\left(  C\right)  }^{ab})\ .\label{pilambft}%
\end{equation}
Using cumbersome but simpe calculations, one can present the finite
transformations (\ref{exply}) for the generating functional $Z_{F}(J)$ in
(\ref{z(0)}) for the model under consideration with the quantum action $S$
given by (\ref{S(FT)}). At the same time, for the purpose of connecting the
integrand $\mathcal{I}_{\Gamma}^{\left(  {F_{0}+\Delta F}\right)  }$ of
$Z_{F_{0}+\Delta F}(J)$ given by a gauge $F_{0}+\Delta F$ with the one given
by a gauge $F_{0}$, so that $\mathcal{I}_{\Gamma}^{\left(  {F_{0}+\Delta
F}\right)  }=\mathcal{I}_{\Gamma}^{\left(  {F_{0}}\right)  }$, as suggested in
Discussion below, it is sufficient, due to the solution of the compensation
equation (\ref{eqexpl}), to find the explicit form of $\lambda_{a}\left(
\phi,\pi,\lambda|\Delta{F}\right)  $ in (\ref{solcompeq2}). To this end, let
us consider a finite change of the gauge condition:%
\begin{equation}
\Delta{F}=F(\alpha,\beta,\gamma)-F_{0}=\int d^{4}x\,\left(  -\frac
{\alpha-\alpha_{0}}{4}\mathfrak{B}_{\mu\nu}\mathfrak{B}^{\mu\nu}-\frac
{\beta-\beta_{0}}{2}\varepsilon_{ab}C_{\mu}^{a}C^{\mu b}-\frac{\gamma
-\gamma_{0}}{12}\varepsilon_{ab}\varepsilon_{cd}C^{ac}C^{bd}\right)
\ .\label{deltaFft}%
\end{equation}
The corresponding field-dependent BRST-antiBRST transformation (\ref{exply})
which provide the coincidence of the vacuum functionals, $Z_{F_{0}+\Delta
F}=Z_{F_{0}}$, are determined by the functionally-dependent odd-valued
parameters:
\begin{equation}
\lambda_{a}\left(  \phi,\pi,\lambda|\Delta{F}\right)  =-\frac{1}{2i\hbar}%
\sum_{n=1}\frac{1}{n!}\left[  \frac{1}{4i\hbar}\Delta{F}\overleftarrow{s}%
^{2}\right]  ^{n}\left(  \Delta{F}\overleftarrow{s}_{a}\right)
\ .\label{funcdeplafinFT}%
\end{equation}

\section{Discussion}

\label{Concl}

\setcounter{equation}{0} In the present work, we have proved that the finite
BRST-antiBRST transformations for a general gauge theory in Lagrangian
formalism announced in \cite{MRnew} are actually invariance transformations
for the integrand in the path integral $Z_{F}(0)$, given by (\ref{z(0)}). To
this end, we have explicitly calculated the Jacobian (\ref{ansjacob})
corresponding to the given change of variables with constant parameters
$\lambda_{a}$. Using the finite BRST-antiBRST transformations, we have
obtained the Ward identity (\ref{WIsp2}) depending on constant parameters
$\lambda_{a}$. The identity contains the usual $\mathrm{Sp}(2)$-doublet of
Ward identities, as well as a new Ward identity at the second order in powers
of $\lambda_{a}$. We have illustrated the construction of finite BRST-antiBRST
transformations in general gauge theories by the example of a reducible gauge
model of a non-Abelian antisymmetric tensor field \cite{FT}.

In conclusion, note that the structure of finite BRST-antiBRST transformations
with field-dependent parameters,%
\begin{equation}
\Delta\Gamma^{p}=\Gamma^{p}\left(  \overleftarrow{s}{}^{a}\lambda_{a}+\frac
{1}{4}\overleftarrow{s}{}^{2}\lambda^{2}\right)  \ ,\ \ \ \lambda_{a}%
=s_{a}\Lambda\ ,\ \ \ \Lambda=\Lambda\left(  \phi,\pi,\lambda\right)
\ ,\label{Gamma_finfd}%
\end{equation}
is the same as in the case of finite field-dependent BRST-antiBRST
transformations in the Lagrangian formalism for Yang--Mills theories
\cite{MRnew}, as well as in the case of the generalized Hamiltonian formalism
\cite{MRnew1}. Consequently, it is natural to expect that the Jacobian
corresponding to this change of variables with functionally-dependent (due to
$s^{1}\lambda_{1}+s^{2}\lambda_{2}=-s^{2}\Lambda$) parameters, inspired by the
infinitesimal field-dependent BRST-antiBRST transformations of
\cite{MRnew,BLT1,BLT2}, should have the form\footnote{The representation for
the Jacobian (\ref{superJaux}), (\ref{superJ1}) has been recently proved in
\cite{MRnew3}.}%
\begin{align}
&  \exp\left(  \Im\right)  =\exp\left[  -\left(  \Delta^{a}S\right)
\lambda_{a}-\frac{1}{4}\left(  \Delta^{a}S\right)  \overleftarrow{s}%
_{a}\lambda^{2}\right]  \exp\left[  \ln\left(  1+f\right)  ^{-2}\right]
\ ,\ \ \mathrm{with}\ \ \,f=-\frac{1}{2}\Lambda\overleftarrow{s}{}%
^{2},\label{superJaux}\\
&  d\check{\Gamma}=d\Gamma\ \exp\left[  \frac{i}{\hbar}\left(  -i\hbar
\Im\right)  \right]  =d\Gamma\ \exp\left\{  \frac{i}{\hbar}\left[  {i}{\hbar
}\left(  \Delta^{a}S\right)  \lambda_{a}+\frac{i\hbar}{4}\left(  \Delta
^{a}S\right)  \overleftarrow{s}_{a}\lambda^{2}+i\hbar\,\mathrm{\ln}\left(
1-\frac{1}{2}\Lambda\overleftarrow{s}{}^{2}\right)  ^{2}\right]  \right\}
\ .\label{superJ1}%
\end{align}
Here, $\Lambda\left(  \phi,\pi,\lambda\right)  $ is a certain even-valued
potential with a vanishing ghost number, and the integration measure $d\Gamma$
transforms with respect to the change of variables $\Gamma\rightarrow
\check{\Gamma}=\Gamma+\Delta\Gamma$ given by (\ref{Gamma_finfd}). Hence,
a compensation equation required to satisfy the relation%
\begin{equation}
Z_{F+\Delta F}=Z_{F}\ ,\label{zphizphi1}%
\end{equation}
as one subjects $Z_{F+\Delta F}$ to a change of variables $\Gamma
^{p}\rightarrow\check{\Gamma}{}^{p}$, according to (\ref{Gamma_finfd}), has
the form%
\begin{equation}
i\hbar\,\mathrm{\ln}\left(  1-\frac{1}{2}\Lambda\overleftarrow{s}{}%
^{2}\right)  ^{2}=-\frac{1}{2}\Delta F\overleftarrow{s}{}^{2}%
\ \Longleftrightarrow\ \left(  1-\frac{1}{2}\Lambda\overleftarrow{s}{}%
^{2}\right)  ^{2}=\exp{\left(  \frac{i}{2\hbar}\Delta F\overleftarrow{s}{}%
^{2}\right)  \ },\label{eqexpl}%
\end{equation}
or, equivalently,
\begin{equation}
\frac{1}{2}\Lambda\overleftarrow{s}{}^{2}=1-\exp\left(  {{\frac{1}{\,4i\hbar}%
}\Delta F\overleftarrow{s}{}^{2}}\right)  \ .\label{eqexpllam}%
\end{equation}
The solution of this equation for an unknown Bosonic functional $\Lambda
\left(  \phi,\pi,\lambda\right)  $, which determines $\lambda_{a}\left(
\phi,\pi,\lambda\right)  $ in accordance with $\lambda_{a}=\Lambda
\overleftarrow{s}_{a}$, with accuracy up to BRST-antiBRST exact ($s^{a}$ being
restricted to $\phi$, $\pi_{a}$, $\lambda$) terms, is given by%
\begin{equation}
\Lambda(\Gamma|\Delta{F})=\frac{1}{2i\hbar}g(y)\Delta{F}\ ,\ \ \mathrm{for}%
\ \ g(y)=\left[  1-\exp(y)\right]  /y\ \ \mathrm{and}\ \ y\equiv\frac
{1}{4i\hbar}\Delta F\overleftarrow{s}{}^{2}\ ,\label{solcompeq2}%
\end{equation}
whence the corresponding field-dependent parameters have the form%
\begin{equation}
\lambda_{a}\left(  \Gamma|\Delta{F}\right)  =\frac{1}{2i\hbar}g(y)\left(
\Delta{F}\overleftarrow{s}_{a}\right)  \ .\label{funcdeplafin}%
\end{equation}
Making in (\ref{z(0)}) a field-dependent BRST-antiBRST transformation
(\ref{Gamma_finfd}) and using the relations (\ref{aexv}) and (\ref{superJ1}),
one can obtain a \emph{modified Ward} (\emph{Slavnov--Taylor})
\emph{identity:}%
\begin{equation}
\left\langle \left\{  1+\frac{i}{\hbar}J_{A}\phi^{A}\left[  \overleftarrow
{s}^{a}\lambda_{a}(\Lambda)+\frac{1}{4}\overleftarrow{s}^{2}\lambda
^{2}(\Lambda)\right]  -\frac{1}{4}\left(  \frac{i}{\hbar}\right)  {}^{2}%
J_{A}\phi^{A}\overleftarrow{s}^{a}J_{B}(\phi^{B})\overleftarrow{s}_{a}%
\lambda^{2}(\Lambda)\right\}  \left(  1-\frac{1}{2}\Lambda\overleftarrow
{s}^{2}\right)  ^{-2}\right\rangle _{F,J}=1\ .\label{mWI}%
\end{equation}
Due to the presence of $\Lambda(\Gamma)$, which implies $\lambda_{a}(\Lambda
)$, the modified Ward identity depends on a choice of the gauge Boson
$F(\phi)$ for non-vanishing $J_{A}$, according to (\ref{solcompeq2}),
(\ref{funcdeplafin}). Notice that the corresponding Ward identities for
Green's functions, obtained by differentiating (\ref{mWI}) with respect to the
sources, contain the functionals $\lambda_{a}(\Lambda)$ and their derivatives
as weight functionals. The Ward identities are readily established due to
(\ref{mWI}) for constant $\lambda_{a}$ in the form (\ref{WIlag1}),
(\ref{WIlag2}). Finally, (\ref{mWI}), with account taken of
(\ref{funcdeplafin}), implies the following equation, which describes the
gauge dependence for a finite change of the gauge $F\rightarrow F+\Delta F$:%
\begin{align}
Z_{F+\Delta F}(J) &  =Z_{F}(J)\left\{  1+\left\langle \frac{i}{\hbar}J_{A}%
\phi^{A}\left[  \overleftarrow{s}^{a}\lambda_{a}\left(  \Gamma|-\Delta
{F}\right)  +\frac{1}{4}\overleftarrow{s}^{2}\lambda^{2}\left(  \Gamma
|-\Delta{F}\right)  \right]  \right.  \right.  \nonumber\\
&  -\left.  \left.  (-1)^{\varepsilon_{B}}\left(  \frac{i}{2\hbar}\right)
^{2}J_{B}J_{A}\left(  \phi^{A}\overleftarrow{s}{}^{a}\right)  \left(  \phi
^{B}\overleftarrow{s}_{a}\right)  \lambda^{2}\left(  \Gamma|-\Delta{F}\right)
\right\rangle _{F,J}\right\}  \;,\label{GDI}%
\end{align}
thereby extending (\ref{zphizphi1}) to the case of non-vanishing $J_{A}$.
Note, that we have proved our conjecture as to the representation
(\ref{superJaux}), (\ref{superJ1}) of the Jacobian for field-dependent
BRST-antiBRST transformations with functionally-dependent parameters in
\cite{MRnew3}.

We have shown, on the basis of field-dependent BRST-antiBRST transformations,
the way to reach an arbitrary gauge, determined by a quadratic (in fileds)
gauge Boson (\ref{F(FT)}) for the Freedman--Townsend model in the path
integral representation, starting from the reference frame with a gauge Boson
$F_{0}$ and using finite field-dependent BRST-antiBRST transformations with
the parameters $\lambda_{a}\left(  \phi,\pi,\lambda|\Delta{F}\right)  $ given
by (\ref{funcdeplafinFT}).

\section*{Acknowledgments}

A.R. is thankful to D. Bykov, D. Francia and to participants of International
Seminar "QUARKS 2014" for useful comments and discussions. The study was
carried out within the Tomsk State University Competitiveness Improvement
Program and was supported by RFBR grant under Project No. 12-02-000121 and by
the grant of Leading Scientific Schools of the Russian Federation under
Project No. 88.2014.2.

\end{document}